\documentclass[prl,showpacs,twocolumn,aps,superscriptaddress,a4paper]{revtex4-1}
\usepackage{dcolumn}
\usepackage{amsmath}
\usepackage{graphicx}
\usepackage{latexsym}
\usepackage{amsfonts}
\usepackage{amssymb}
\usepackage{color}
\usepackage{pstricks,pst-node,pst-text,pst-3d,graphpap,pst-plot}

\DeclareGraphicsExtensions{.eps,.pdf,.gif,.jpg,.png}

\def\g{\gamma}
\def\G{\Gamma}

\def\e{\epsilon}

\newcommand{\be}{\begin{equation}}
\newcommand{\ee}{\end{equation}}
\newcommand{\beq}{\begin{eqnarray}}
\newcommand{\eeq}{\end{eqnarray}}

\begin{document}

\title{Insight on Hole-Hole Interaction and Magnetic Order from
Dichroic Auger-Photoelectron  Coincidence Spectra}

\author{M. Cini}
\affiliation{Dipartimento di Fisica, Universit\`a
di Roma Tor Vergata, Via della Ricerca Scientifica 1, I-00133 Rome,
Italy}
\affiliation{Laboratori Nazionali di Frascati, Istituto Nazionale di
Fisica
Nucleare, Via E. Fermi 40, 00044 Frascati, Italy}

\author{E. Perfetto}
\affiliation{Dipartimento di Fisica, Universit\`a
di Roma Tor Vergata, Via della Ricerca Scientifica 1, I-00133 Rome,
Italy}

\author{R. Gotter}
\affiliation{IOM-CNR Istituto Officina dei Materiali del CNR, Area
Science Park - Basovizza, Trieste, Italy}

\author{G. Offi}
\affiliation{CNISM and Dipartimento di Fisica, Universita' di Roma
Tre-Rome,Italy}

\author{A. Ruocco}
\affiliation{CNISM and Dipartimento di Fisica, Universita' di Roma
Tre-Rome,Italy}

\author{G. Stefani}
\affiliation{CNISM and Dipartimento di Fisica, Universita' di Roma
Tre-Rome,Italy}

\begin{abstract}

The absence of sharp structures in the core-valence-valence Auger
line shapes of partially filled bands  has severely  limited
the use of electron spectroscopy in magnetic crystals and other correlated materials.
Here  by a
novel interplay of experimental and theoretical techniques we achieve 
 a combined understanding of the Photoelectron,
Auger 
and Auger-Photoelectron Coincidence
Spectra (APECS) of CoO. This is a prototype antiferromagnetic material in which
the recently discovered Dichroic Effect in Angle Resolved (DEAR) APECS reveals a complex
pattern in the strongly correlated Auger line shape.
A  calculation of the \textit{unrelaxed} spectral features explains
the pattern in detail, labeling the final states by the total spin.
The present theoretical analysis shows that the dichroic  effect
arises from a spin-dependence of the  angular distribution of the photoelectron-Auger electron
pair detected in coincidence, and from  the selective power
of the dichroic  technique in assigning different weights to the
various spin components. Since the spin-dependence
of the  angular distribution exists in the antiferromagnetic state but vanishes at the N\'eel
temperature, the DEAR-APECS technique detects the phase transition from
its local effects, thus  providing  a unique  tool to
observe and understand magnetic correlations in such circumstances,  where the usual methods (neutron diffraction, specific heat measurements) are not applicable.

\end{abstract}


\maketitle

The Auger Core-Valence-Valence (CVV) transitions, that produce a pair
of holes in the valence states of an atom in a solid, have long been
of interest. Powell\cite{powell}
pointed out that some crystals  have  broad band-like CVV profiles
while others show  atomic-like multiplet  spectra.
In the Cini-Sawatzky
theory\cite{resonances1,resonances2, sawatzky}, which,
with some refinements, yields accurate   line
shapes\cite{cole} without free  parameters\cite{fratesi}, the line shape gives
details of the  on-site dynamics and screened interactions in solids.
Strong correlations produce two-hole resonances and atomic-like spectra,
while weak correlations produce a distortion of the band self-convolution.
The short range of the Auger
matrix elements (a few atomic units) ensures that the information is
\textit{local} (that is, available also from tiny  inhomogeneous samples)
and microscopic in character.
Early work was limited to closed (i.e. fully occupied)
valence bands.  An extension\cite{ldaauger} to open bands with
an occupation of at least $\sim 90\%$ led to the
Bare-Ladder Approximation\cite{lda,palladio,clusters}. This theory
provides  accurate
results for photoemission and Auger spectra in cases like Pd
metal\cite{clusters}.
However at lower filling the open  bands polarize strongly
around the primary core hole. Then, the dynamics is more involved
since it cannot be described in terms of the propagation of two holes,
and the theoretical analysis
complicates substantially\cite{gs,cinidrchal,marini,ugenti,perfetto,nanotubes}.
In a simplified approach proposed by  Drchal and one of us\cite{cinidrchal},
some spectral features are \textit{unrelaxed}, i.e., arise from the propagation
of two holes in a unpolarized rigid  background, similar to the closed-bands case;
the \textit{relaxed} part of the spectrum comes instead from a screened situation
and involves the screening cloud. Anyhow for decades, little information has been gained from open-band spectra.
Indeed, the Auger $L_{23}VV$ line shape from  metals like Cr,  Fe, Co and compounds
like CoO are almost featureless\cite{australiani}, characterized by broad structures, eventually with
little Coster-Kronig satellites\cite{coster}.
However, despite the spectra have a band-like look, the magnetic properties
of these solids indicate that correlations must be quite strong.

The APECS (Auger-Photoelectron Coincidence Spectroscopy) technique measures Auger spectra originated by  a specific core-hole state.  Important input came from the further  discovery\cite{haak,stefani-review,dear}  that the coincidence Auger electron energy distribution depends on the emission angle of  both Auger electron and Photoelectron.  This was called Dichroic Effect or DEAR  APECS. It was argued that spin-symmetric (i.e., high-spin) and spin-antisymmetric (i.e., low-spin) final states are enhanced or suppressed depending
upon the chosen geometry, but this lacked theoretical support.

In this  Letter  we show that substantial progress in the
understanding of magnetic correlations in open bands solids can be
achieved by the interplay between theory and the DEAR-APECS technique.
We model the CoO crystal by an octahedral
CoO$_{6}$ cluster centered on the  Co ion. We include a minimal basis set and
fix the parameters of the model by comparing the calculated Local
Density of States (LDOS) with the experimental XPS profile.
Then, we show
how it is possible to reveal considerable structure in the Auger data,
and interpret it in terms of the spin selectivity of APECS. Finally we point
out that this case study has general implications about electron
spectroscopy and also in the field of magnetism.

The linear combinations of $3d$ orbitals which form the $e_{g}$ and
$t_{2g}$ irreducible representations (irreps) of the Octahedral Group
are a suitable one electron basis for the Co ion.
We  take into account only the combinations of
$2p$ orbitals with symmetry $e_{g}$ and $t_{2g}$
\be
p_{\G\g}=\sum_{\substack{J\\ j=x,y,z}}c_{Jj}^{\G\g}\,p_{Jj},
\ee
where $J$ runs over the 6 Oxygens, $\G$ runs over the irreps and $\g$
over the corresponding components.  In this basis
the non-interacting part of the Hamiltonian reads
\beq
H_{0}=\sum_{\G=e_{g},t_{2g}}\sum_{\g}\left[
\e_{d}(\G)d_{\G\g}^{\dagger}d_{\G\g}+\e_{p}(\G)p_{\G\g}^{\dagger}p_{\G\g}\right.
\nonumber\\ \left.
+t_{pd}(\G,\g)(d_{\G\g}^{\dagger}p_{\G\g}+p_{\G\g}^{\dagger}d_{\G\g}
)\right]\eeq
where  we take $\e_{d}(e_{g})= \Delta-6D$, $\e_{p}(e_{g})=-(\epsilon_{\sigma}-\epsilon_{\pi})$
and $\e_{d}(t_{2g})= \Delta+4D$, $\e_{p}(t_{2g})=\epsilon_{\sigma}-\epsilon_{\pi}$.
The hopping parameters are the following linear combinations of Slater-Koster matrix elements
\cite{sk.1954} $E_{\G\g,j}(J)$:
\be
t_{pd}(\G,\g)=\sum_{\substack{J\\
j=x,y,z}}c_{Jj}^{\G\g}E_{\G\g,j}(J).
\ee
In the above Equation the Oxygen $J$ is specified by the direction cosines
$(l,m,n)$ and the energies $E_{\G\g,j}(J)$ are expressed in terms of $\tau_{\sigma}$ and
$\tau_{\pi}$ transfer integrals. As in previous works\cite{resonances1,resonances2} we
include the on-site repulsion in the Co ion  and neglect it  on the O
sites.  The interaction part of the Hamiltonian  is taken in the standard form:
\be
H_{\rm int}=\sum_{mm'nn'}\sum_{\sigma \sigma^{\prime}}U_{mm'nn'}
d^{\dagger}_{m\sigma}d^{\dagger}_{m'\sigma^{\prime}}
d_{n^{\prime}\sigma^{\prime}}d_{n\sigma},
\ee
where the  $m,m',n,n'$ run over  the Co orbitals; the  $U_{mm'nn'}$
elements can be written  in terms  of
Slater integrals which, in turn,  are expressed
in terms of the Racah parameters\cite{racah} $A$, $B$, $C$ according
to $f^{4}=\frac{63}{5}C$, $f^{0}=A+\frac{1}{9}f^{4}$,
$f^{2}=\frac{441}{9}B+\frac{5}{9}f^{4}$.
As only $A$ is affected by the solid state screening we use $B=0.14$ eV and
$C=0.54$ eV
like in the isolated Co\cite{racah}. Eventually  by a unitary transformation
we rewrite $H_{\rm int}$  in the symmetry adapted basis $(\G,\g) $.

\begin{figure}[]
\includegraphics*[width=.23\textwidth]{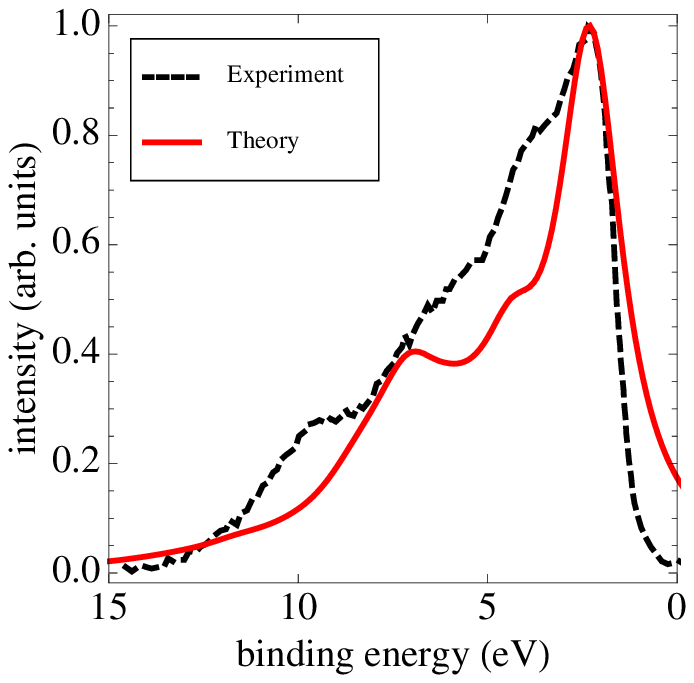}
\includegraphics*[width=.23\textwidth]{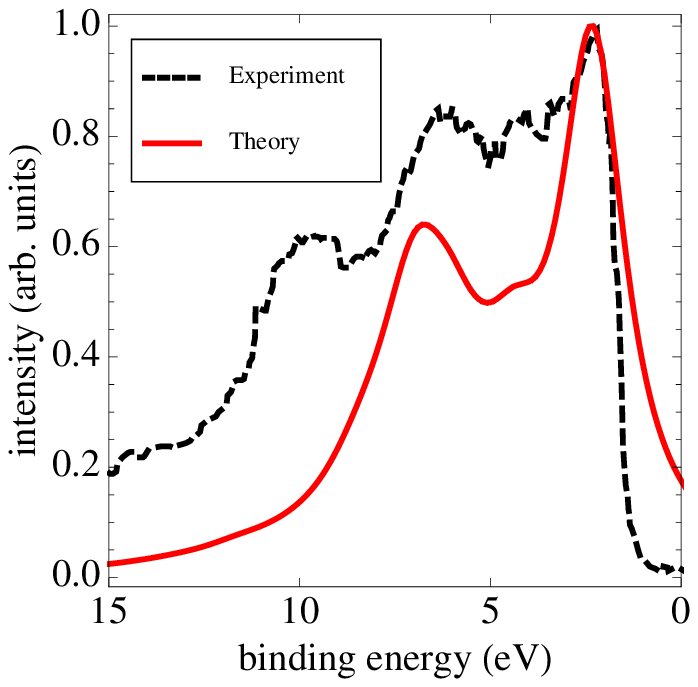}
\caption{Comparison of experimental and calculated valence-band  XPS profiles.
Left panel: the spectrum excited by  1253.6 eV X-rays from
Ref. \onlinecite{vanelp} (dashed line) and our
calculated Co-removal spectrum (solid line).  Right panel:
 the spectrum excited by  250 eV X-rays
from Ref. \onlinecite{collapse} (dashed line) and our calculated ($0.8
\mathrm{Co}+
0.2 \mathrm{O}$)-removal spectrum (solid line). In the calculated
spectra a 1 eV Lorentzian broadening was used.
}
\label{xps}
\end{figure}

In the left panel of Fig. \ref{xps} we reproduce from
Ref. \onlinecite{vanelp} the XPS spectrum obtained by an
unmonochromated Mg K$\alpha$ line. Using the
above geometry and basis set, van Elp and coworkers  also computed the Co LDOS\cite{vanelp}  by a
continued fraction technique (excluding some  higher energy
configurations)  with a Lorentzian  broadening of 1 eV.  The 3-holes ground state
($d^7$ configuration) was taken
to belong to  maximum spin  $S=3/2$ in agreement with Hund's
rule. Choosing  the parameters that gave a visual agreement  a with their XPS spectrum,
 they estimated  $A=5.2$ eV.
In the right panel of Fig. \ref{xps}
we also report the profile obtained by soft X-Rays
 from Ref. \onlinecite{collapse}. The lower energy
of the incident photons gives rise to a rather different
spectrum. The difference can be understood qualitatively
because the ionization cross section for $p$ states drops with
frequency faster than that of the more localized $d$ states, and
so soft X-Rays data have much stronger oxygen character.

The exact diagonalization of CoO$_{6}$ model Hamiltonian with 3 holes
shows that  the ground state with the
parameters of Ref. \onlinecite{vanelp} is at energy -11.78 eV for spin 3/2
and -13.03 eV for spin 1/2.
Thus Hund's rule does {\it not} apply unless $A$ is considerably
increased with the consequence, however, of deteriorating the
agreement with experimental profile.
We interpret this result as an artifact of the
small cluster, which reduces the degeneracy of one-electron levels
forcing the electrons to arrange in a low-spin  configuration.
Therefore, we consider the $S=3/2$ sector
as the most appropriate one for  modeling the
actual solid. Taking  (in eV) $\Delta=5.5, D=0.07, \epsilon_{\sigma}=0.55$
and $ \epsilon_{\pi}=-0.15$ as in Ref.\onlinecite{vanelp} we diagonalised
the Hamiltonian with 4 holes and
computed the LDOS at the Co and O sites,
looking for the best Racah $A.$
The priority for the theory is to reproduce the
positions of the most prominent features  at lower binding energies,
which are similar in both spectra.
To  ensure  that the prominent narrow peak near
the Fermi level has a dominant Co character in both cases,
and correctly produce the enhancement of the structure
(with dominant O character) at 7.4 eV in the right panel, we need  $A=1.08$,
which is definitely smaller  than the value found in Ref. \cite{vanelp}.
In Fig. \ref{xps} we display the pure Co
profile (left panel, solid curve) and, according to the above
discussion, the one with  20\% Oxygen component (right panel, solid curve).

To conclude the  part on XPS  we stress that the  spectrum  does not
change significantly as the temperature is increased
from $T=170$ K (in the
AF phase) to $T=295$ K (above the N\'eel
temperature).

\begin{figure}
\includegraphics*[bb=7 65 242 335, width=.38\textwidth]{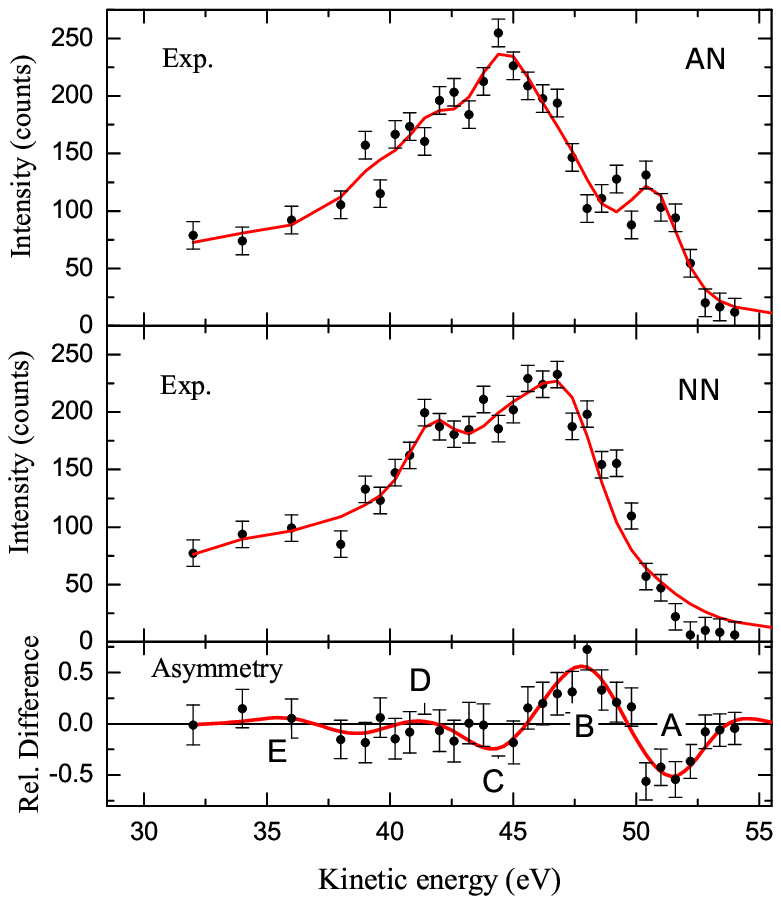}\\
\includegraphics*[width=.38\textwidth]{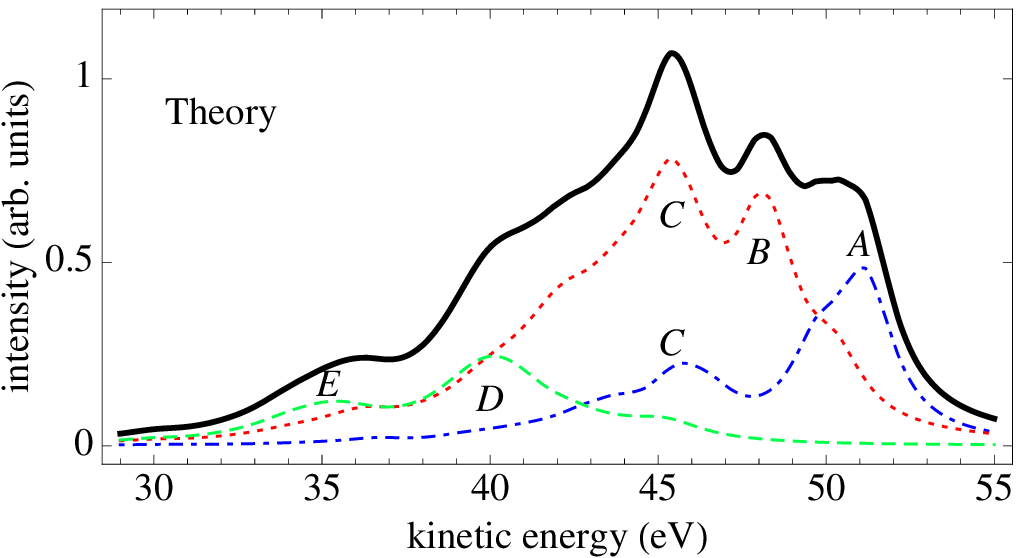}
\caption{(Color online) Top panel: Experimental DEAR-APECS spectra of CoO {\it vs}
Auger electron kinetic energy for the AN
 and NN geometry, and their difference
(dashed).
The continuous AN  and NN lines are obtained by a best fit procedures
based on a representation of the experimental  line shapes by six Voigt
profiles (see main text). The difference (or asymmetry curve) displays
structures labeled by $A-E$.
Bottom panel: Calculated total Auger spectrum (solid line) and the different
contributions corresponding to the spin $S=5/2$ (dot-dashed)  $S=3/2$
(dotted)
$S=1/2$ (dashed) of the final state.  The structures $A-E$ are reproduced
by the theory. In addition, $C$ has a shoulder at $\sim 43$ eV which is
not evident in the asymmetry curve because it appears in both the $S=5/2$ and
$S=3/2$ spectra, but is clearly seen in both the $AN$ and $NN$ data.
}
\label{augerspectrum}
\end{figure}


The experimental $M_{23}M_{45}M_{45}$  Auger spectra of CoO from
Ref. \onlinecite{collapse} (and reported in
the upper panels of Fig. \ref{augerspectrum}) were  measured
in coincidence with  the  photoelectron current excited from the Co
$3p_{\frac{3}{2}}$ core level.  The profiles marked AN and NN
are APECS spectra taken at different angles,
at a low temperature, i.e., with the CoO film
in the AF  phase. The notation  NN means
that both electrons are not aligned (N) with the photon polarization
whereas  AN means that the photoelectron is aligned with the photon polarization
while the Auger electron is not.
It was observed\cite{collapse} that the low- spin
two-hole final states are favored in the NN  geometry while  the AN configuration
favors high-spin (DEAR APECS effect).
By comparing  AN and NN  spectra, a clear modulation of the line
shape is obtained and sharp peaks emerge. In particular the asymmetry curve (i.e., the difference between NN and AN spectra) 
displays a fingerprint, which cannot be seen in the \textit{singles} spectrum.
The most remarkable point is that effect is lost above the
N\'eel temperature\cite{collapse} while XPS and the singles Auger
lines do not change at the transition.


\begin{center}
\begin{table}
$\begin{tabular}{|c|c|c|c|c|c|}
  \hline
  Feature  & A & B & C & D & E \\
  \hline
  Spin &$5/2$&$3/2$&$3/2,5/2$&$1/2$&$1/2$\\
  Theory (eV) & 51.5& 48.0 & 45.4& 40.3 & 35.5 \\
  Experiment  (eV) & 50.7 & 47.2 & 44.6 & 39.5 & 34.7 \\
  Intensity AN & 285& 377& 1020& 301& 361\\
  Intensity NN & 64& 785& 819& 308& 365\\
  \hline
\end{tabular}
 $\\
\caption{Main features in the APECS spectrum of CoO, characterized
by their kinetic
energies, relative intensities and main spin components. The experimental
energies and intensities  are obtained by the best fit described in the text.}
 \end{table}
\end{center}

We below address this issue and calculate
the Auger the unrelaxed spectrum\cite{cinidrchal} without any adjustable parameter, keeping the
values optimized for the XPS lineshape. The $M_{45}$ final-state holes belong to the
Co$3d$-O$2p$ valence band.
Up to a proportionality constant, the Auger current
$J(\epsilon_{k})$ with wavevector $k$
after the absorption of a photon of energy $ \omega$ by a core
level $c$, and
accompanied by a photoelectron of energy $\epsilon_{p}$ is
\be
J(\epsilon_{k})=\sum_{\alpha,\beta,\gamma,\delta}
M_{ck\alpha\beta}^{*}M_{ck\gamma\delta}
D_{\alpha\beta\delta\gamma}(\epsilon_{p}+\epsilon_{k}- \omega),
\label{spectrum}
\ee
where $-\pi D_{\alpha\beta\delta\gamma}(\omega)$ is
the imaginary part of the Fourier transform of the retarded
 the Green's function
\be
G_{\alpha\beta\gamma\delta}(t)=
-\langle T\left[d_{\beta}^{\dagger}(t)d_{\alpha}^{\dagger}
(t)d_{\gamma}d_{\delta} \right]\rangle,
\ee
and $M_{ck \alpha \beta}$ is the  Auger matrix element, obtained
by the Coulomb integral between the core state  $c$, the
Auger electron $k$,
and the two valence states labelled by $\alpha$ and $\beta$.
We computed the  $M_{ck\alpha\beta}$  matrix elements for the
$M_{23}M_{45}M_{45}$  transitions using the Clementi-Roetti
atomic orbitals \cite{clero} for positive ions and  plane-waves  for the Auger
electrons;  only the diagonal contributions $(\alpha,\beta)=(\gamma, \delta)$
in Eq. (\ref{spectrum})
were included and  the direction of the momentum $k$ was taken along the quantization axis of
the Co ion. Such a simple treatment is reasonable since
we are not aiming at absolute
rate calculations but at computing the line shape.
A Lorentzian  of width  1 eV was convolved  to the function $J$ to
simulate lifetime and any other broadening effects.

The CoO$_{6}$ cluster  (10  orbitals)  hosts  5  holes after the Auger
decay and the rank  of the problem is
$\left(\!\!
\begin{array}{c}
20 \\
5 \\
\end{array}
\!\!\right)=
15504$.
 Let $H_{nm}$ denote the Hamiltonians with $m$
spin-up and $n$ spin-down holes;  we put
$H_{50},H_{41}$ and $H_{32}$  in block form using the total spin $S$
symmetry and found  the eigenvectors of the blocks. $H_{32}$ has 252 sextets,
1848 quartets and 3300 doublets which form the biggest block.
In the bottom panel of Fig. \ref{augerspectrum} we report the calculated,
spin-resolved  Auger spectrum. The dot-dashed, dotted and dashed  lines show
the  contribution of $S=5/2$,  $S=3/2$  and
$S=1/2$ respectively, while the solid line is the total angular unresolved spectrum.
The letters $A-E$ are used to mark the main features whose maxima in
energy are reported in Table I.

To make comparison of theory and experiment, we need to know the
positions of the experimental peaks. To this end, we represented
the AN and NN spectra by superpositions of six Voigt profiles and made
partial  best fits.
The results are shown in Table I.
Taking into account that in the experiment of Ref. \onlinecite{collapse}
no absolute calibration of the Auger energy scale was performed
and that  an overall uncertainty of $\pm$0.4 eV is estimated,
the energy position of the five main structures correspond very
well with the theoretical predictions for both geometries. 

%

The present theoretical analysis demonstrates that the DEAR-APECS
technique detects structure in the line shape because the AN and NN
geometries discriminate the  spin components; $A$ which is mainly
$S=5/2$ prevails in AN while $B$ which is mainly $S=3/2$
weights more in NN. Qualitatively, the different spin contents of
structures A and B had been conjectured already  in Ref. \onlinecite{collapse}
based on a Tanabe-Sugano
\cite{tanabe} analysis of the data.
The present more quantitative work validates this picture and shows that
the same occurs at lower kinetic energies too  (features $D$ and $E$)
where a similar dichroism occurs between
$S=1/2$ and $S=3/2$.  The best fit intensities reported in Tab. I show
that the NN geometry generally favors high spin states of the final ion and
the AN geometry prefers low spin; in general terms, the APECS geometry
strongly affects the probability of leaving the Co  ion in different spin states.

Equation (\ref{spectrum}) can capture only the {\em unrelaxed}
spectrum and  is already sufficient to assign the main  structures.
The {\em relaxed} spectrum must be responsible for the broad and
rather flat background that one sees in the experimental profile, but
{\it a posteriori} we conclude that it does not play a very important role in this case. This may be due to
a short $3p$ lifetime  (the energy width of the $3p$ photoemission line is found to be 1.1 eV)
or to slow screening of the core-hole.


In conclusion,
by a tight interplay of theory and experiment we succeeded to
observe, identify  and characterize in some detail the final states of a
core-valence-valence transition in a correlated open band.
This finding solves a
long standing problem and is an important achievement in the field of
electron spectroscopy.
Previously, such spectra had been  misinterpreted as  band-like despite the
fact that the materials had magnetic properties.
The spin selectivity is inherent in the dichroic technique, since
the ion spin governs the angular distribution
of the photoelectron-Auger electron pair, thus allowing to monitor the
magnetism at the ion site.   We must  tackle new exciting
problems, including a full  theory of the operation of
DEAR-APECS.  An extended theory should explain the disappearance of
the effect in the paramagnetic phase, that is, why  not only $S_{z}$ but also
$S$   fluctuates during the lifetime of the core-hole fast enough
to kill the effect. In this way, we will be able to  better
understand  and exploit its implications as a local probe
of the magnetic order in solids.

\section{Acknowledgment}
Financial support from the MIUR PRIN 2008 contract
prot. 2008AKZSXY is gratefully acknowledged.

\end{document}